\documentclass[fleqn,10pt]{wlscirep}
\usepackage[utf8]{inputenc}
\usepackage[T1]{fontenc}
\title{Phase transitions in the mini-batch size for sparse and dense two-layer neural networks}

\author[1,2,*]{Raffaele Marino}
\author[1,3,+]{Federico Ricci-Tersenghi}

\affil[1]{Sapienza Università di Roma, Dipartimento di Fisica, Piazzale Aldo Moro 5, 00185, Roma, Italy}
\affil[2]{Università degli studi di Firenze, Dipartimento di Fisica e Astronomia, Via Giovanni Sansone 1, 50019, Sesto Fiorentino, Firenze, Italy}
\affil[3]{CNR-Nanotec and INFN, sezione di Roma1, Piazzale Aldo Moro 5, 00185, Roma, Italy}

\affil[*]{raffaele.marino@unifi.it}
\affil[+]{federico.ricci@uniroma1.it}

\begin{abstract}
The use of mini-batches of data in training artificial neural networks is nowadays very common. Despite its broad usage, theories explaining quantitatively how large or small the optimal mini-batch size should be are missing. This work presents a systematic attempt at understanding the role of the mini-batch size in training two-layer neural networks.
Working in the teacher-student scenario, with a sparse teacher, and focusing on tasks of different complexity, we quantify the effects of changing the mini-batch size $m$.
We find that often the generalization performances of the student strongly depend on $m$ and may undergo sharp phase transitions at a critical value $m_c$, such that for $m<m_c$ the training process fails, while for $m>m_c$ the student learns perfectly or generalizes very well the teacher. Phase transitions are induced by collective phenomena firstly discovered in statistical mechanics and later observed in many fields of science. Observing a phase transition by varying the mini-batch size across different architectures raises several questions about the role of this hyperparameter in the neural network learning process.
\end{abstract}
\begin{document}

\flushbottom
\maketitle

\thispagestyle{empty}

\section{Introduction}

The widespread diffusion of neural networks in many scientific fields urges a better understanding of the processes that underlie their training. The discipline that studies how well simple devices like Turing machines or more abstract constructions, such as classifier systems, can learn and infer from observed data after the training process, is the statistical learning theory \cite{hastie2009elements}. The statistical learning theory is at the cornerstone of Machine Learning (ML), and it deals with the statistical inference problem of finding a predictive function based on data. Even though neural networks are well analyzed from statistical learning, they have become an active field of research in statistical mechanics. Statistical mechanics predicts the properties of a macroscopic system from the laws of its microscopic dynamics \cite{huang2008statistical, baldovin2023ergodic, marino2016entropy}. In this area, a major role is played by phase transitions that regulate what is achievable in principle (information theoretical thresholds) and what is achievable in practice (algorithmic thresholds)\cite{zdeborova2016statistical,coja2017information,caracciolo2021criticality, banks2016information, malatesta2019fluctuations, capelli2018exact, malatesta2017two, marino2016backtracking, coja2020information}. The graphical representation (the phase diagram) of the various phases delimited by phase transitions allows researchers to predict the response of the system as a function of its own tunable parameters.  The phase diagrams of simple but realistic neural networks ( e.g., perceptron, one hidden layer, committee machine \cite{franz2019critical,amit1990retrieval,sclocchi2021proliferation}) are well known and understood, when the algorithmic dynamics is governed by equilibrium processes \cite{engel2001statistical}. However, the out-of-equilibrium dynamics of the learning processes \cite{agoritsas2018out,marino2023hard, decelle2021equilibrium} are much more difficult to study and most of the results about it are restricted to dense models where the Martin-Siggia-Rose formalism \cite{PhysRevA.8.423} can be applied to derive DMFT equations \cite{opper2001advanced, gerbelot2022rigorous}.

Statistical mechanics tools have been also applied in the realm of artificial intelligence \cite{lecun2015deep, goodfellow2016deep} for building up consistent theories for deep learning \cite{xu2019explainable, baldassi2021unveiling, baldassi2022learning,lucibello2022deep, pittorino2017chaos}.
%
A deep neural network is a type of machine learning model, and when a deep network is fitted to data, this is referred to as deep learning \cite{prince2023understanding}.
Deep learning (DL) has shown very powerful empirical performance for solving very complex real-world problems in areas such as computer vision \cite{krizhevsky2012imagenet}, natural language processing \cite{collobert2011natural, cabessa2021efficient}, speech recognition \cite{yu2016automatic}, recommendation systems \cite{wei2017collaborative}, drug discovery \cite{hutson2019ai}, differential equations \cite{shloof2023new,marino2023solving}, and much more \cite{silver2016mastering,marino2021learning, shah2022study}.
%
In simple words, deep learning can be seen a neural network \cite{bollobas1998modern}, composed by many layers, that takes some data set $\mathcal{D}$, input and targets, and learns the rules for forecasting new input data.
For learning the rules, one minimizes a loss function by optimizing the weights of the neural network, using an optimization algorithm \cite{rumelhart1986learning, lecun2012efficient, kingma2014adam, pittorino2021entropic, hastings1970monte, kirkpatrick1983optimization,MarinoAurell2016, aurell2016diffusion, earl2005parallel}. The weights are collected into a set $\Theta$.
The peculiarity of a deep learning model is the application of a non-linear function on each output of the hidden layer, and, in general, on each neuron of the output layer. These non-linear functions are called \textit{activation functions}.
Statistically, deep learning estimates a function $\hat{f}(\vec{x},\Theta)$, with $\vec{x}$ the input data, that minimizes a loss function $\mathcal{L}(\vec{y}, \hat{f}(\vec{x},\Theta))$, where $\vec{y}$ represents the target. This minimization is performed over a set of pairs $(\vec{x}, \vec{y})_{\eta}$, where $\eta$ indices an element of $\mathcal{D}$ and $|\mathcal{D}|$, the cardinality of $\mathcal{D}$,  can be finite or infinite.

Over the last decades, the practitioners of neural networks have developed many very useful tricks and smart procedures, like mini-batch \cite{li2014efficient}, dropout\cite{srivastava2014dropout}, 
%
deep residual learning framework (feedforward neural networks with “shortcut connections”) \cite{he2016deep} 
and several other regularizations \cite{prince2023understanding} for speeding up the training process. 

A theory justifying many of these choices is often lacking, and so it is very difficult to make optimal choices for who is not an expert practitioner. Among these "tricks" the use of the so-called mini-batch, introduced as a technical requirement for dealing with huge databases, actually turns out to be crucial for optimal training.

In machine learning, a mini-batch is a subset of the full dataset that is used to train a model \cite{li2014efficient}. Rather than training the model on the entire dataset at once, the training data is divided into smaller batches, or mini-batches, which are fed to the model one at a time. Mini-batch training is a commonly used optimization technique in deep learning. It enables the model to make multiple updates to its weights and biases based on the gradients computed from each mini-batch. This process of making small updates to the model weights is called Stochastic Gradient Descent (SGD) \cite{bottou2012stochastic, chaudhari2019entropy, lecun1998gradient}. The size of the mini-batch is a hyperparameter that can be tuned to optimize the training process. A larger mini-batch size can lead to faster training times, but it can also make it harder for the model to converge to a good solution. A smaller mini-batch size can improve the model's convergence but may slow down the training process. Mini-batch training provides a compromise between the two extremes of batch training (using the entire dataset at once) and online training (updating the model after each individual data point).

This hyperparameter is chosen for fitting the GPU hardware during the training process, and only few theoretical analysis have been performed to understand if exists or not an optimal value for it \cite{perrone2019optimal, masters2018revisiting,  smith2017don, smith2018disciplined}.  
In Ref.~\citeonline{perrone2019optimal}, the authors present an empirical law for selecting the mini-batch size that minimizes Stochastic Gradient Descent learning time for single and multiple learner problems, where the empirical law depends by some empirical parameters that must be deduced by the data, the model topology and the learning algorithm used. In other words, the empirical law is not universal. In Ref.~\citeonline{masters2018revisiting}, the authors review the small mini-batch methods for deep learning and  present numerical analyses in which the best performance is obtained for mini-batch sizes between $m = 2$ and $m = 32$. The results provided evidence that increasing the mini-batch size implies a degradation of the test performance and a progressively smaller range of learning rates that allows stable training. However, there is no theoretical motivation supporting these results. The manuscripts Ref.~\citeonline{smith2017don} and Ref.~\citeonline{smith2018disciplined} showed that one can improve the performance of Stochastic Gradient Descent, on both training and test sets, by keeping constant the learning rate and increasing the mini-batch size during training. Those results were obtained over different architectures of neural networks.

From the point of view of statistical physics, understanding the key role of the mini-batch in the learning process, in particular, whether it exists a phase transition that sets an optimal value for the mini-batch size, is still an open question.
Here we consider this problem and provide a positive answer by showing the existence of a phase transition in the mini-batch size ruling the ability to train optimally neural networks. More precisely, we confine ourselves  into the "Teacher-Student" scenario, where the \textit{teacher} is sparse and with binary weights, while the \textit{student} can be sparse, or dense, with binary or continuous weights depending on the information that the teacher gives to the student.
%
We choose this setting because it is simple and we can keep control on the true model that we can infer or generalize. 
The teacher builds its own neural network and creates an infinite set of data $\mathcal{D}$. 
She gives to the student the whole data set and asks to build her own  neural network. The student can have information or not about the topology of the teacher network. 
In the case she knows everything, she needs only to infer the sign of the teacher's weights. Therefore, the student could use a greedy algorithm. If she does not know the topology of the neural network, she is allowed to build a neural network with continuous weights and use a Stochastic Gradient Descent algorithm for generalizing the teacher weights.

The "Teacher-Student" model has been a useful tool for building up knowledge in science, and in deep learning. For instance, in Ref.~\citeonline{goldt2019dynamics} the authors studied the generalization analysing the dynamics and the performance of over-parametrized two-layer neural networks in the teacher-student setup, showing that the dynamics of the SGD, in the online learning case, can be described by a set of differential equations that are asymptotically exact in the limit of large inputs, in the case where the student networks have more parameters than the teacher. In Ref.~\citeonline{cornacchia2022learning }, instead, the authors provide closed-form asymptotic expressions for multi-class teacher-student perceptron generalization errors in the high-dimensional regime. Again, in the high-dimensional regime some authors \cite{ loureiro2021learning} studied the generalisation of the model where the teacher and student can act on different spaces, generated with fixed, but generic feature maps. They show a rigorous formula for the asymptotic training loss and generalisation error.

In this setup, the few analytical results on the existence of phase transitions in the thermodynamic limit are for the Ising perceptron \cite{engel2001statistical}.  In this case, the generalization error as a function of the ratio between the number of training examples and the dimension of the input space has a first-order phase transition that allows identifying the value at which the Ising perceptron learns.

These analyses, however, leave open the issue of whether it exists a phase transition for the mini-batch size in two-layer neural networks.
For filling this gap, in this paper, we present, as far as we know for the first time, the existence of a phase transition in the hyperparameter $m$ such that, for each value of $m$ smaller than $m_c$, the critical value of mini-batch size,  the generalization/inference of the teacher weights is impossible, while for each value of $m$ bigger than $m_c$ the generalization/inference is possible. Moreover, we present strong evidence that the phase transitions are independent of the model used or the algorithm used, and that they seem to be of the first order, depending on the value of the sparsity in the teacher model. To make our results general and robust we study several neural network topologies and inference problems.
We will consider both dense and sparse topologies. The latter is much less used, but there are promising future applications given the large saving in memory and computing power.



The paper is organized as follows: in Sec. \ref{sec:summary} we make a short summary of our results.  In Sec. \ref{sec:model} we introduce the probabilistic model called "Teacher-Student", the notation used, and the algorithms used for the numerical analysis. In Sec. \ref{sec::numanal} we present our numerical analysis for the models listed in Sec. \ref{sec:model}. In Sec. \ref{Conc} we discuss the results.

\section{Short Summary of the Main Results}
\label{sec:summary}
To help the reader of getting the main message of this manuscript, here we summarize our results. 

We study four different teacher-student models. 
\begin{itemize}
    \item The first one, i.e., \textbf{ST-SS-2M-OC} (see Sec. \ref{sec:model} and Sec. \ref{mod1}), the teacher and the student have the same sparse topology embedded by a bipartite random $d$-regular graph between the input and the hidden layer, and between the hidden and the multidimensional output layer, with LeakyReLu activation functions on the hidden and output neurons. In this case, we observe that the normalized validation loss goes to zero discontinuously for each value of $d > 2$ at different values of the mini-batch size $m$. In other words, the student easily can infer the weights of the teacher only above a particular value of the mini-batch size $m_c$.
    \item In the second model, i.e., \textbf{ST-SS-2M-MC} (see Sec. \ref{sec:model} and Sec. \ref{mod2}), the teacher is the same as the first one, while the student neural network is embedded with a random $d_s$-regular graph, with $d_s$ bigger than the one of the teacher $d_t$ (the activation functions are LeakyReLu). In other words, the aim of the student is to infer the weights of the teacher and at the same time understand which weights of its own network must be set to zero for reaching the perfect inference. We observe the existence of phase transitions in the mini-batch size, but this time of two different natures. For $d_t=2$ and $d_s>2$ we observe a second-order phase transition, while for $d_t>2$ and $d_s>d_t$ we observe again phase transitions of the first order.
    \item The third model, i.e., \textbf{ST-SS-1M-OC} (see Sec. \ref{sec:model} and Sec. \ref{mod3}), differs from the other two by changing the activation function, which now becomes an $erf(\cdot) $ on the hidden neurons and identity on the output neuron. The multidimensional output layer now becomes one-dimensional. The student knows the topology of the teacher network and needs only to infer the weights. In this case, the perfect inference is not anymore possible, however, as shown in Fig.~\ref{fig:loss_reg_1l}, the existence of a first-order phase transition in the size of the mini-batch is still present. This time, the transition divides two regions, one where the generalization is poor, and another where it is possible having good performance in generalizing the planted teacher weights.
    \item The fourth model, i.e., \textbf{ST-DS-1M-MC} (see Sec. \ref{sec:model} and Sec. \ref{mod4}), has a teacher network which is the same as in the case of  \textbf{ST-SS-1M-OC}, but the student does not know anything about the topology of the network. She has access only to the data and creates her own dense neural network.  By tuning the hyperparameters of the student model, we observe that the presence of an optimal mini-batch that allows a very good generalization is still present in correspondence with another optimal value of the learning rate (associated with the Stochastic Gradient Descent used as an optimizer). This model will be analysed more deeply in future work, but it suggests a possible existence of phase transitions.
\end{itemize}

\section{The Models and the Algorithms}
\label{sec:model}

In this section, we introduce the probabilistic model called "Teacher-Student". This model is pretty simple. We have one teacher and one student. The teacher builds a neural network, and asks the student to infer its weights, given a data set $\mathcal{D}$ made of input-output pairs.

In other words, a teacher generates a random neural network. She generates a certain number $M$ of input vectors $\vec{x}_{\eta}$,  $\eta=1,\ldots,M$ and computes the associated outputs using the neural network, i.e., $\vec{y}_{\eta}$,  $\eta=1,\ldots,M$, where $M$ is $|\mathcal{D}|$. The student, thus, is provided with the data, i.e. the input-output pairs $(\vec{x}_{\eta},\vec{y}_{\eta})_{\eta \in [M]}$, and her objective is to infer teacher's weights from these data. 
The teacher could give or not the exact topology of the model to the student: if she does, then, we call this scenario optimal-case, otherwise, we call it mismatch-case.  

For example, consider a supervised regression task. The data set $\mathcal{D}$ is composed of $M$ pairs $(\vec{x}_{\eta}, \vec{y}_{\eta})_{\eta \in [M]} \in \mathbb{R}^{d_{x}+d_{y}}$ identically and independently sampled from $\mathbf{P}(\vec{x}, \vec{y})$. The prior probability $\mathbf{P}(\vec{x})$ is assumed to be known and $\mathbf{P}(\vec{y}|\vec{x})$ is modeled by a two layer neural network. 
Given a feature vector $\vec{x}_{\eta} \in \mathbb{R}^{d_x}$, the respective label $\vec{y}_{\eta} \in \mathbb{R}^{d_{y}}$ is defined as
\begin{equation}
\label{eq::model}
\vec{y}_\eta = \psi\left(\mathbf{W}^*_\text{out}\,\phi\left(\mathbf{W}^*_\text{in}\,\vec{x}_\eta\right)\right),
\end{equation}
where $\phi: \mathbb{R} \to \mathbb{R}$ and $\psi: \mathbb{R} \to \mathbb{R}$ are two activation functions that act element-wise, while $\mathbf{W}^*_\text{in}$ is a $k \times d_x$  matrix and $\mathbf{W}^*_\text{out}$ is a $d_{y} \times k$ matrix. In most cases, we choose these matrices to be sparse, with non-zero elements taking values $\pm 1$, 

Given a new sample $\vec{x} \sim \mathbf{P}(\vec{x})$ outside the training data, the goal is  to obtain an estimation function $\hat{f}(\vec{x}, \Theta): \mathbb{R}^{d_x} \to \mathbb{R}^{d_y}$ (where $\Theta$ is an arbitrary set of parameters to be learned from the data) for the respective label $\vec{y}$. The error is quantified by a loss function $\mathcal{L}(\vec{y}, \hat{f}(\vec{x}, \Theta))$. The loss function used in this manuscript is a quadratic loss of the type
\begin{equation}
\label{eq::loss}
    \mathcal{L}(\vec{y}, \hat{f}(\vec{x}, \Theta)) = \frac{1}{2} \sum_{\eta = 1}^M (\hat{f}(\vec{x}_{\eta}, \Theta) - \vec{y}_{\eta} )^2
\end{equation}

We are interested in understanding the role of sparsity in neural networks.  More precisely, our goal is to estimate the teacher model with another two-layer neural network with the same activation functions and the same number of neurons in each layer, which we will refer to as the student.  Formally the student model reads
\begin{equation}
\label{eq::studentmodel}
     \hat{f}(\vec{x}_{\eta}, \Theta) = \psi\left(\mathbf{W}_\text{out}\,\phi\left(\mathbf{W}_\text{in} \vec{x}_\eta\right)\right),
\end{equation}
where $\Theta$ identifies the set of the parameters that must be inferred, i.e., the elements of the matrices $\mathbf{W}_{\text{out}}$ and $\mathbf{W}_{\text{in}}$.

\subsection{Training Algorithms}

We allow the student to choose between two different algorithms for estimating the teacher model. More precisely, the student can choose between a greedy algorithm and a Stochastic Gradient Descent algorithm (to be better described below), depending on the nature of the parameters to be optimized, which in turn depends on the information provided by the teacher.
We choose to work with a virtually infinite data set, i.e.\ we never present to the student twice the same data item, because we have found a very weak dependence of the results on the data set size $M$ and a fast convergence towards the large $M$ limit.
Once the algorithm is fixed and the data set infinitely large, the most relevant parameter in the training process is the mini-batch size $m$, that is how many data items are used in each step of the training.  We will show that an optimal choice for this parameter can drive the algorithms to infer/generalize the teacher model. To the best of our knowledge, this is the first time a phase transition in this important parameter is found.

The first algorithm that the student can use for minimizing the loss function is a greedy algorithm. This algorithm is used when the teacher provides the student with information about the discrete nature of the matrix elements.
The greedy algorithm is just a Metropolis updating rule \cite{metropolis1953equation} at zero temperature, that is the proposed new $\Theta$ configuration must imply a lower value for the loss function (computed on the validating set) in order to be accepted.
At zero temperature there are no thermal fluctuations helping the evolution to escape local minima in the loss function. However, the finite (and small) value of $m$ introduces a different kind of noise that can play the same beneficial role in escaping local minima to reach the optimal configuration. 

At any given step of the greedy algorithm, an element $w_{ij}$ of one of the matrices entering the model definition \eqref{eq::studentmodel} is selected at random and let us assume the support of this weight is $\pm 1$. Calling $\overline{w}_{ij}=-w_{ij}$ the new value we propose for this weight, the change in loss is given by
\begin{equation}
\label{diff::loss}
\Delta \mathcal{L} = \frac{1}{2} \sum_{\eta = 1}^m (\hat{f}\left(\vec{x}_{\eta}, ((\Theta \setminus \{w_{ij}\}) \cup \{\overline{w}_{ij}\})) - \vec{y}_{\eta}\right)^2 - \frac{1}{2} \sum_{\eta = 1}^m (\hat{f}\left(\vec{x}_{\eta}, \Theta) - \vec{y}_{\eta} \right)^2 .
\end{equation}
%

We define with  $\Theta \setminus \{w_{ij}\}$ the set of parameter $\Theta$ where the element $w_{ij}$ has been removed. Therefore, $(\Theta \setminus \{w_{ij}\}) \cup \{\overline{w}_{ij}\}$ defines the set of parameter $\Theta$ where element $w_{ij}$ has been substituted with the element $\overline{w}_{ij}$.
The algorithm accepts the flip only if $\Delta \mathcal{L}\le 0$. This difference $\Delta \mathcal{L}$ depends on the mini-batch size $m$ and can fluctuate a lot changing the mini-batch for small values of $m$. 
We use the same mini-batch set for an entire Monte Carlo Sweep, i.e.\ the attempt to update all the non-zero matrix elements. 
%

During this manuscript, we will also analyze a model where the support of the random variable is given by $\{1,-1,0\}$. In such a case, we modify slightly the Metropolis update, for fitting the new support of the random variable, by sampling uniformly the new state where the weights should go.

The second algorithm is a Stochastic Gradient Descent algorithm. This algorithm is used when no information is given to the student about the teacher model. Thus, the student can model the teacher by her choice and chooses to use real variables for $\Theta$. To update the weights the student computes the gradient of the loss function in \eqref{eq::loss} over a mini-batch of size $m$ and uses the usual updating rule
\begin{equation}
\label{SGDupdate}
\Theta \leftarrow \Theta -{\lambda}\nabla \mathcal{L}(\vec{y}, \hat{f}(\vec{x}, \Theta), m),
\end{equation}
where $\nabla$ is the gradient operator over the weights and $\lambda$ is the learning rate. 

\subsection{Teacher-student scenarios}

The above algorithms are used for analyzing different cases of the teacher-student scenario. We list them below. 

\begin{itemize}
    
    \item \textbf{Sparse teacher-student, two matrices, optimal-case (ST-SS-2M-OC)}. In this case, the teacher is sparse and the matrices $\mathbf{W}^{*}_{\text{in}}$ and $\mathbf{W}^{*}_{\text{out}}$ have non-zero values equal to ${\pm 1}$. The student knows the topology of the neural network, i.e., she knows the position of the non-zero values in $\mathbf{W}^{*}_{\text{in}/\text{out}}$, and she knows that the non-zero elements of these matrices can take only the values $\pm 1$. The student needs only to infer  the sign of the weights. The analysis of this model is explained in Sec.~\ref{mod1}.
    
    \item \textbf{Sparse teacher-student, two matrices, mismatch-case (ST-SS-2M-MC)}. In this case, the teacher is sparse and the matrices $\mathbf{W}^{*}_{\text{in}}$ and $\mathbf{W}^{*}_{\text{out}}$ have non-zero values equal to $\pm 1$. The student does not know exactly the topology of the neural network, as the teacher provides her with a topology of larger connectivity, that contains the teacher network. The student is allowed to infer the teacher network by setting some weights to zero. So the student weights will take values in $\{\pm 1, 0\}$. The analysis of this model is explained in Sec.~\ref{mod2}
    
    \item \textbf{Sparse teacher-student, one matrix, optimal-case (ST-SS-1M-OC)}. In this case, the teacher matrix $\mathbf{W}^*_\text{in}$ is still sparse with $\pm 1$ elements and needs to be inferred, while the matrix $\mathbf{W}^*_\text{out}$ is fixed and provided to the student. For simplicity, we consider a scalar output $y$ (i.e.\ $d_{y}=1$) equal to the mean of the neurons of the hidden layer. The student knows also the topology of the teacher's neural network and that weights take values in ${\pm 1}$. So, she just needs to infer the sign of the weights. The analysis of the model is explained in Sec.~\ref{mod3}
     
    \item \textbf{Sparse teacher, dense student, one matrix, mismatch-case (ST-DS-1M-MC)}. As in the previous case, the teacher matrix $\mathbf{W}^{*}_{\text{in}}$ is sparse, with $\pm 1$ weights, while the $\mathbf{W}^{*}_{\text{out}}$ is dense with all elements equal to $1$ (i.e.\ takes the mean of the hidden layer). The student is provided with the matrix $\mathbf{W}^{*}_{\text{out}}$ and needs to infer the matrix $\mathbf{W}^{*}_{\text{in}}$. Having no information about the topology of the latter, she decides to train a dense neural network with the aim of generalizing the values of the sparse matrix $\mathbf{W}^{*}_{\text{in}}$ with continuous values. The analysis of the model is explained in Sec.~\ref{mod4}
\end{itemize}

All this model will be fully explained in the next section, where we will also present the numerical results.

\section{Results}

\label{sec::numanal}
In this section, we present our results on each single model listed in Sec. \ref{sec:model}. 

\subsection{Sparse teacher-student, two matrices, optimal-case}
\label{mod1}

For this analysis, we assume the framework given in \eqref{eq::model}. The teacher builds a neural network with the same amount of neurons in each layer. More precisely, the teacher builds $\mathbf{W}^{*}_{\text{in}}$ and $\mathbf{W}^{*}_{\text{out}}$ to be sparse, with $d_x = d_y = k = N$. The total number of non-zero entries is $Nd$ for each matrix. In other words $\mathbf{W}^{*}_{\text{in}}$ and $\mathbf{W}^{*}_{\text{out}}$ are two square matrices $N \times N$, where each row and each column contains only $d$ non-zero elements (in positions randomly chosen). The activation functions $\psi$ and $\phi$ are non linear and set to be $\text{LeakyRELU}(x)$:

\begin{equation}
\label{eq:leakyrelu}
\text{LeakyRELU}(x) = \left\{
        \begin{array}{ll}
            ax & \quad x < 0 \\
            x & \quad x \geq 0
        \end{array}
    \right.
\end{equation}

We fix the parameter $a=0.1$. In this analysis, the student knows the model and the positions of the non-zero entries in the two matrices. She also knows that non-zero weights are discrete, taking values $\pm 1$. The activation functions are known as well. The student builds her own neural network and decides to use a greedy algorithm for inferring the values of the matrices elements. She defines the loss function as in \eqref{eq::loss}. 

\begin{figure}
    \centering
    \includegraphics[width=0.7\textwidth]{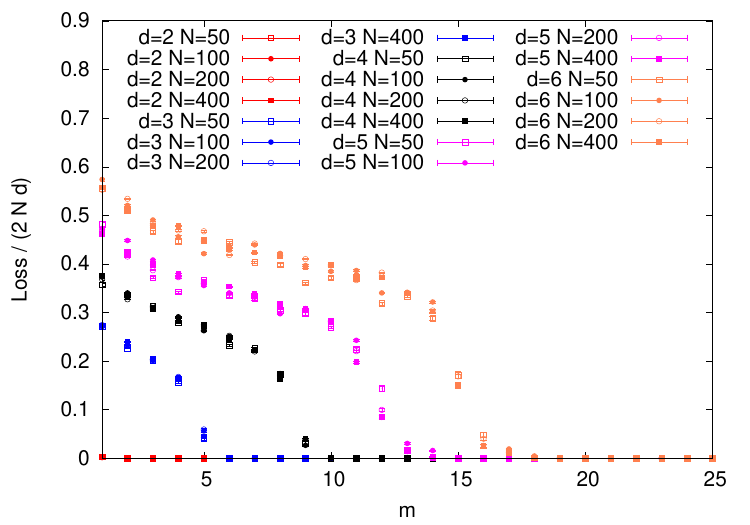}
    \caption{The figure shows the validation loss function normalized to the number of non-zero elements, averaged over $100$ samples, as a function of $m$, i.e., the mini-batch size, for different values of $d$, and different values of $N$. Error bars are standard deviation of the mean.}
    \label{fig:MLnormvsm_1}
\end{figure}

Under the above assumptions, we perform an accurate analysis of the model as follows. We build the teacher neural network, fixing the value of $N$ to $50, \,100,\,200,\,400$ and the parameter $d$, i.e., the number of non-zero entries in each row and column of matrices $\mathbf{W}^*_{\text{in/out}}$, to $d=2,\,3,\,4,\,5,$ and $6$. Each non-zero element is $\pm 1$ with $50\%$ probability. We sample the independent components of the input vectors $\vec{x} \in \mathbb{R}^N$ from a normal distribution of zero mean and unitary variance, i.e.\ $x_i \sim \mathcal{N}(0, 1)$.  Thus, we create the associated labels $\vec{y} \in \mathbb{R}^N$, using the neural network defined above.

Given the teacher neural network defined by $\mathbf{W}^*_{\text{in/out}}$, the two weight matrices $\mathbf{W}_{\text{in/out}}$ defining the student network have non-zero elements in the same positions, and only the signs need to be inferred.

For every value of $N$ and $d$, we perform the numerical analysis, starting from a random initialization for the student weights and aiming at minimizing the loss in \eqref{eq::loss} using the greedy algorithm described above with different mini-batch size $m$. Each Monte Carlo sweep \footnote{One Monte Carlo sweep is equal to $2Nd$ attempts to update a weight.} is performed with a new set of size $m$ of pairs $(\vec{x}_\eta,\vec{y}_\eta)_{\eta=1,\ldots,m}$. We never use twice the same data pair (as in the online learning problem) or equivalently we assume to have an infinite amount of data to train the network with. Moreover, we define a set of pairs $(\vec{x}_{\eta},\vec{y}_{\eta})_{\eta=1,..,30}$ to be the validation data set. These data are used only for computing the validation loss, and they are never used during the training process.  

\begin{figure}
    \centering
    \includegraphics[width=0.5\textwidth]{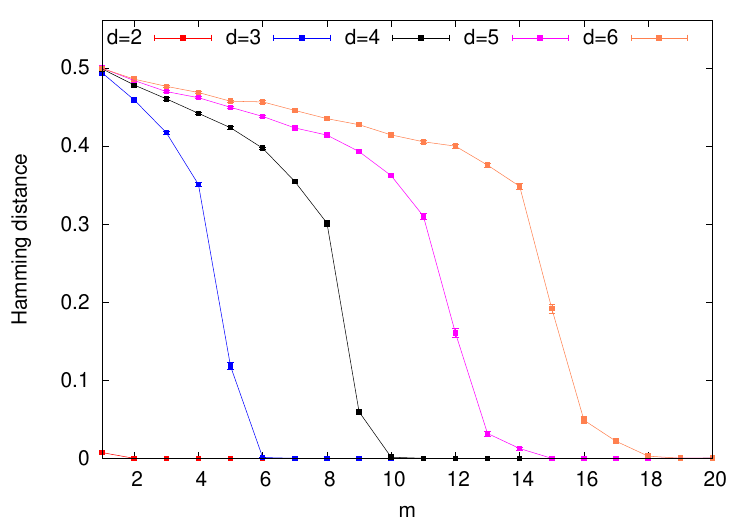}
    \includegraphics[width=0.495\textwidth]{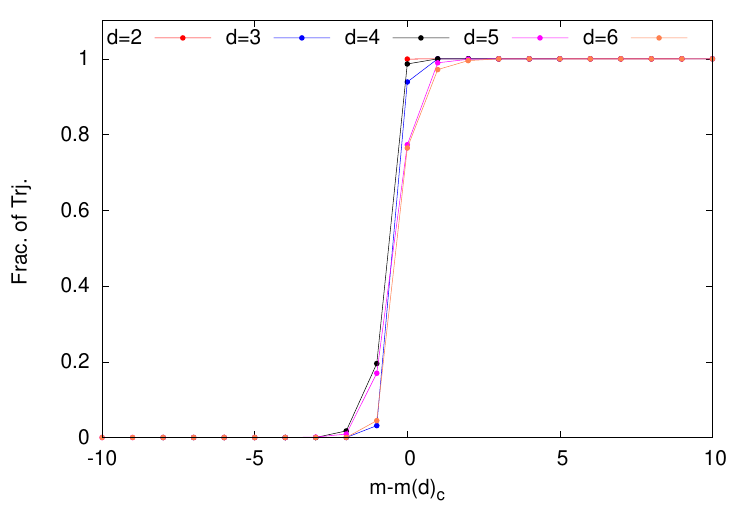}
    \caption{\textbf{Left}: The figure displays the averaged Hamming distance over $100$ samples as a function of the mini-batch size $m$, for different values of the parameter $d$, i.e., the number of non-zero entries in each row and column of each matrix $\mathbf{W}^*_{\text{in/out}}$. Error bars are standard deviation of the mean. \textbf{Right}: The figure displays the fraction of Hamming distance trajectories that have found an asymptotic value smaller than $0.25$  as a function of $m-m_c(d)$. In this case  all the curves collapse one on top of the other, showing a step function.}
    \label{fig:HDDcollaps}
\end{figure}

In Fig.~\ref{fig:MLnormvsm_1}, we show the validation loss function normalized to the number of non-zero elements, averaged over $100$ samples, as a function of $m$, i.e., the mini-batch size. The value of the validation loss is the one obtained after a time of $1024$ Monte Carlo sweeps. We have checked that at that time the process has reached a stationary regime and more iterations would not change the results. The curves drawn, for any values of $d$, seem to be independent of the value of $N$ used. In other words, finite size effects are negligible and we can consider the results as in the thermodynamic limit. 

Results in Fig.~\ref{fig:MLnormvsm_1} show that for $d=2$ the inference of the teacher network is possible for any value of $m$, while the curves with $d>2$ show a more interesting behaviour. For $d>2$ the validation loss value jumps discontinuously to zero at a critical value for the mini-batch size that we call $m_c(d)$. This means that by running the learning algorithm with $m>m_c(d)$ the student is able to infer perfectly the teacher network, while perfect learning is impossible for lower values of $m$.

What happens at $m=m_c(d)$ looks like a phase transition. To the best of our knowledge, such a clear phase transition in the mini-batch size was never observed before.
To better understand the nature of this transition,  we study a different observable, the (normalized) Hamming distance between teacher and student matrices weights. The Hamming distance is simply the number of weights elements which are different. The normalized version we use is just divided by the number of non-zero elements.
In case the student guesses at random the signs of the teacher's weights, the mean Hamming distance is equal to $0.5$, while a perfect inference of the teacher matrix would correspond to a zero Hamming distance.

Data for the mean Hamming distance are reported in the left panel of Fig.~\ref{fig:HDDcollaps}. As for the validation loss discussed above, we observe no $N$-dependence in the data and therefore we average the Hamming distance over all values of $N$. The variation of the mean Hamming distance when the value of $m$ is increased is consistent with a discontinuous phase transition at $m_c(d)$ for $d>2$. We define $m_c(d)$ as the first value of $m$ at given $d$ for which the Hamming distance is compatible with $0$.

To get a quantitative measure of the sharpness of the phase transition at $m_c(d)$ we compute the fraction of trajectories (over different samples and different $N$ values) that have a Hamming distance smaller than 0.25 \footnote{The results do not depend on this threshold value as long as it is chosen in the "gap" visible in the left panel of Fig.~\ref{fig:HDDcollaps}.}.
The nice scaling and the step-like behaviour as a function of $m-m_c(d)$ suggest the transition is very sharp (see right panel in Fig.~\ref{fig:HDDcollaps}).

\begin{figure}
    \centering
    \includegraphics[width=0.7\textwidth]{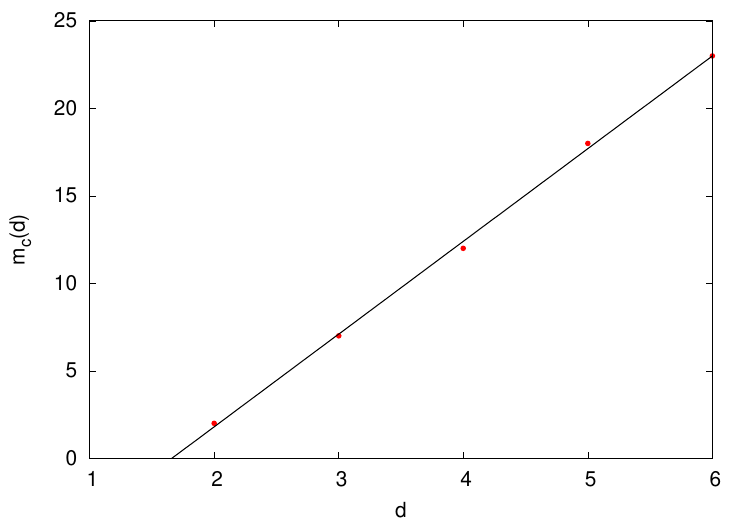}
    \caption{In this figure the value of $m_c$, i.e., the \textit{critical} value of the mini-batch size, as a function of $d$ is plotted. The values of $m_c(d)$ fit well a linear behaviour as a function of $d$.}
    \label{fig:scalingd1}
\end{figure}

The last quantitative aspect to discuss is the dependence of the critical value $m_c$ on the number $d$ of non-zero elements in the network matrices. As shown in Fig. \ref{fig:scalingd1} we find that a linear scaling nicely fit all the data. This is very reasonable: indeed, the number of parameters (i.e.\ signs) to infer is exactly $2dN$ and each mini-batch of data contains $2mN$ numbers. So a linear scaling of $m_c(d)$ seems the best achievable scaling.

\subsection{Sparse teacher-student, two matrices, mismatch-case}
\label{mod2}

For this analysis, we assume the framework given in \eqref{eq::model}, where the teacher is built exactly as in the case discussed in Sec.~\ref{mod1}: matrices $\mathbf{W}^{*}_{\text{in,out}}$ are sparse with exactly $d_t$ non-zero elements per row and column; non-zero weights take values $\pm 1$ with equal probability.
The activation functions $\psi$ and $\phi$ are non-linear and set to be equal to \eqref{eq:leakyrelu}.

In this analysis, the teacher does not give to the student complete information about the topology of her network (i.e.\ the location of non-zero elements in the matrices). The teacher gives to the student just partial information, that is sparse matrices $\mathbf{W}_{\text{in,out}}$ built as before but with $d_s(>d_t)$ non-zero elements per row and column. The network corresponding to this slightly more connected topology is such that it contains the teacher network as a subgraph. In other words, the student can infer the teacher network by just setting some proper weights to zero.
The reason beyond this choice is straightforward: we are interested in checking how general the results shown in Sec.~\ref{mod1} are, in particular, when some sort of noise is added to the inference process.
The present setting is meant to model a situation where a perturbation is introduced in the network topology, such that the ground truth (i.e.\ the teacher network ) is preserved.

The student will work with variables, representing the weights, taking values in $\{\pm1, 0\}$. In such a way she is in principle able to infer the exact network of the teacher if enough information is provided to her.
The student uses the greedy algorithm described in Sec.~\ref{sec:model} to optimize the loss function.
As before, we use an online learning setting, where each data pair $(\vec{x}_{\eta},\vec{y}_{\eta})$ is used for just one Monte Carlo sweep. Moreover 30 data pairs are reserved for the computation of the validation loss and never used in the training.

\begin{figure}
    \centering
    \includegraphics[width=0.49\textwidth]{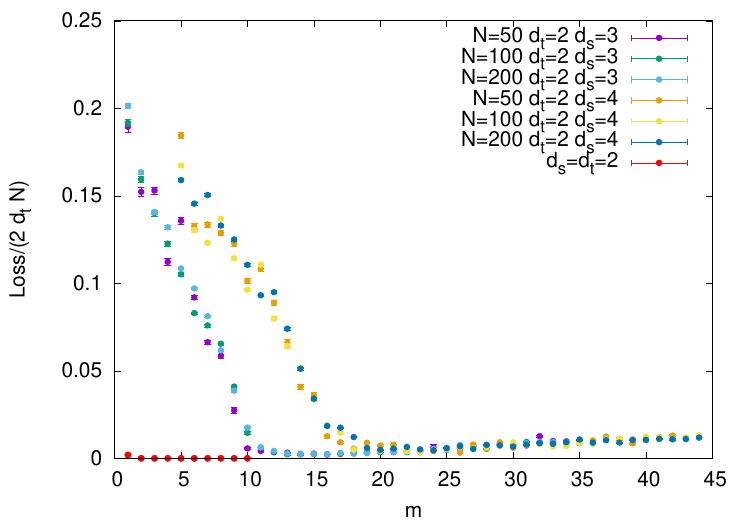}
    \includegraphics[width=0.49\textwidth]{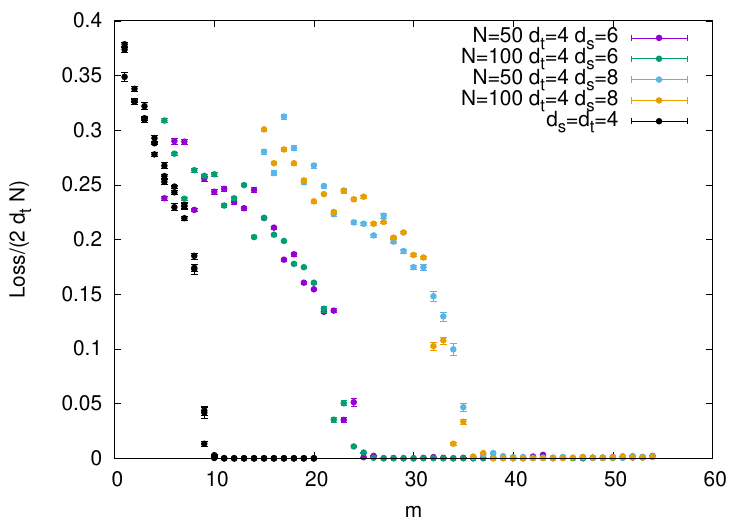}
    \caption{The figures display the averaged validation loss normalized to the number of non-zero elements as a function of the mini-batch size $m$, for different values of $N$ and different values of $d_s$, averaged over $100$ samples (error bars are the standard deviation of the mean). In the left panel we fix $d_t=2$, while in the right panel we fix $d_t=4$.}
    \label{fig:Loss_mod2}
\end{figure}

In Fig.~ \ref{fig:Loss_mod2} we report the mean validation loss function normalized to the number of non-zero elements, averaged over $100$ samples, as a function of $m$, the mini-batch size. In the left panel we have fixed $d_t=2$ and we observe that the curves goes to zero in a continuous way, and without showing any evident finite size effect. We call $m_c(d_s|d_t)$ the critical value of $m$ at which the mean loss attains the zero value. The phase transition at $m_c(d_s|d_t)$ separates a phase where the teacher network can not be correctly learned, from a phase where the student can almost perfectly learn it. A careful inspection of the region $m>m_c(d_s|d_t)$ reveals that the validation loss is not exactly zero, and it increases slightly with $m$. A possible explanation for this observation is the following: the mini-batch size acts as a source of noise that allows the greedy algorithm to better optimize the loss function by escaping from local minima, but if $m$ becomes too large, the effective noise decreases and the training algorithm can get trapped in sub-optimal minima with a very small values of the loss (in any case the  value of the loss is tiny and the student network still generalizes very well the teacher one).

In the left panel of Fig.~\ref{fig:Loss_mod2}, a comparison with the optimal case (i.e.\ $d_s=2$ and $d_t=2$) shows an interesting phenomenon. In the case where the complete information is given to the student, the system does not present any phase transition. Even with a mini-batch of size $m=1$ the student is able to infer almost completely the teacher weights. In contrast, when the teacher hides a little bit of information to the student, providing a noisy version of the network topology, the amount of knowledge that the student needs to infer the teacher weights is much larger. We observe that the minimal size of the mini-batch that the student needs for completely inferring the teacher weights is roughly ten times bigger if $d_s=3$ and even more if $d_s=4$.

In the right panel of Fig.~\ref{fig:Loss_mod2} we present the averaged validation loss obtained by the student when the value $d_t=4$ is used by the teacher. Also in this case, finite size effects are negligible, thus suggesting that we are already probing the large $N$ limit. As before, we compare the optimal case $d_s=d_t=4$ with the mismatched cases $d_s=6,8$. The latter cases require more information to infer the teacher network and this implies a larger critical value $m_c(d_s|d_t)$. The nature of the phase transition at $m_c(d_s|d_t)$ remains discontinuous as in the optimal case, but the jump at the critical point seems to decrease, so we can not exclude that the transition could become continuous for larger values of $d_s$.

\begin{figure}
    \centering
    \includegraphics[width=0.49\textwidth]{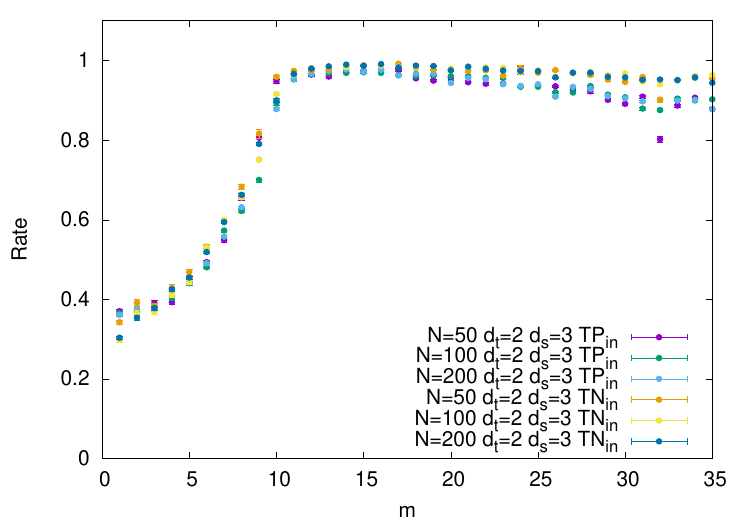}
    \includegraphics[width=0.49\textwidth]{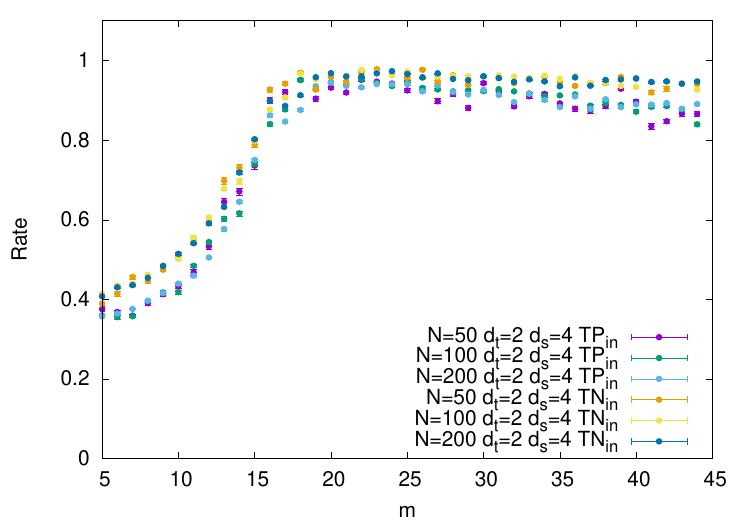}
    \includegraphics[width=0.49\textwidth]{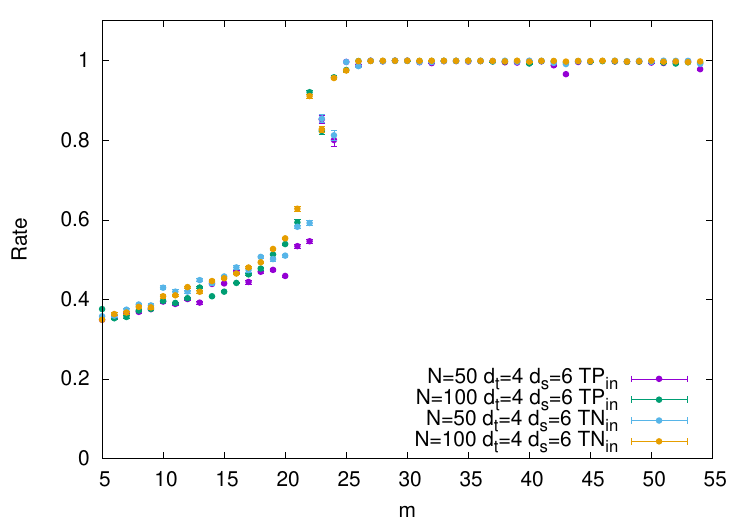}
    \includegraphics[width=0.49\textwidth]{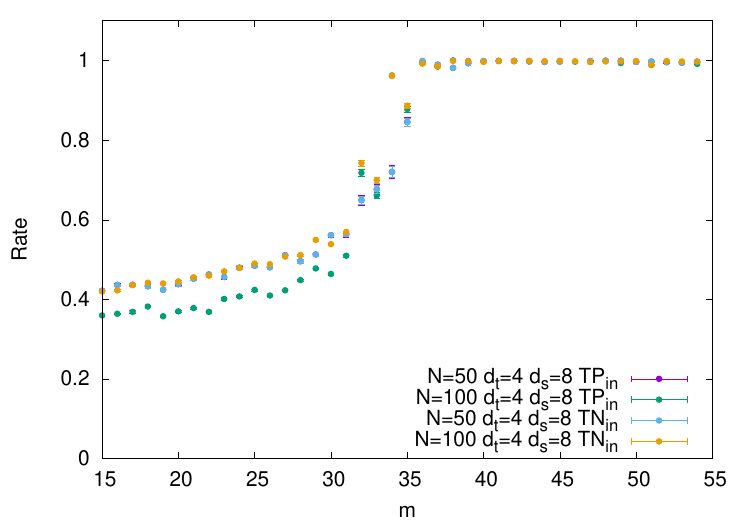}
    \caption{TP and TN rates measure the fraction of correctly inferred non-zero and zero weights in the matrix $\mathbf{W}_\text{in}$. They are plotted as a function of the mini-batch size $m$ for several values of the network connectivity: $d_t=2, d_s=3$ (top left), $d_t=2, d_s=4$ (top right), $d_t=4, d_s=6$ (bottom left), $d_t=4, d_s=8$ (bottom right).
    }
    \label{fig:TFNP}
\end{figure}

The phase transition can be observed also in the fraction of weights inferred correctly by the student. For simplicity, we show in Fig.~\ref{fig:TFNP} results only for the matrix $\mathbf{W}_\text{in}$, but similar results hold for $\mathbf{W}_\text{out}$ as well. We define the True Positive (TP) rate as the fraction of non-zero weights correctly inferred by the students and the True Negative (TN) rate as the fraction of zero weights correctly inferred by the student. In case the teacher network is perfectly recovered, both TP and TN equal 1.

In Fig.~\ref{fig:TFNP}, we show TP and TN rates for different values of $d_t$ and $d_s$. The top panels refer to the $d_t=2$ case, where the TP and TN rates reach the value 1 at $m_c(d_s|d_t=2)$ in a continuous way and slightly decrease for larger $m$ values (in agreement with the results obtained via the validation loss).
The bottom panels are for $d_t=4$ and we observe the TP and TN rates to jump at $m_c(d_s|d_t=4)$ to the value 1, corresponding to a perfect recovery of the teacher network (again in agreement with results from the validation loss).
 
An interesting observation is that increasing the value of $d_s$ for a given $d_t$ has the main effect of translating the transition point, without changing sensibly the nature of the transition. The increase of $m_c(d_s|d_t)$ with $d_s$ is simple to explain: for a larger value of $d_s$ (i.e.\ a larger noise) the number of configurations of the student weights is much greater and much more information is required to select the proper weights.

\subsection{Sparse teacher-student, one matrix, optimal-case}
\label{mod3}

For this analysis, the teacher modifies the setting in \eqref{eq::model} by fixing the matrix $\mathbf{W}^{*}_{\text{out}}$ of size $k \times 1$ and  with all elements equal to $1$.  Moreover, the activation function $\psi$ becomes linear. In practice, the output signal is nothing but the average of the hidden layer. The function $\phi$ remains non-linear, and to make a connection with previous works, we fix $\phi(t)=\text{erf}(t/\sqrt{2})$ and $d_x=k=N$.
Under these assumptions, the teacher model can be written as
\begin{equation}
\label{mod3eq}
    y = \frac{1}{N} \sum_{r=1}^{N}\phi \left( \frac{((\mathbf{W}^{*})^\intercal \vec{x})_r}{\sqrt{d}} \right),
\end{equation}
where $y \in \mathbb{R}$, and $\mathbf{W}^{*}=\mathbf{W}^{*}_{\text{in}}$ to simplify the notation. The matrix $\mathbf{W}^{*}$ is assumed sparse, with exactly $d$ non-zero elements in each row and column, set randomly to $\pm 1$ with equal probability. 

A variant of this model is studied in Ref.~\citeonline{veiga2022phase}, where the authors describe the different regimes of the Stochastic Gradient Descent learning algorithm for this two-layer neural networks in the high-dimensional input layer limit, but with a mini-batch size fixed to $1$. Moreover, in order to solve exactly the dynamics, they need to work with a matrix $\mathbf{W}^{*}$ which is dense for both the teacher and the student. Eventually, the student network is over-parametrized by enlarging the hidden layer with respect to the teacher one ($k_s>k_t$). However, this study provides no clue on what happens changing the mini-batch size, nor making the teacher (and/or the student) sparse.

\begin{figure}
    \centering
    \includegraphics[width=0.49\textwidth]{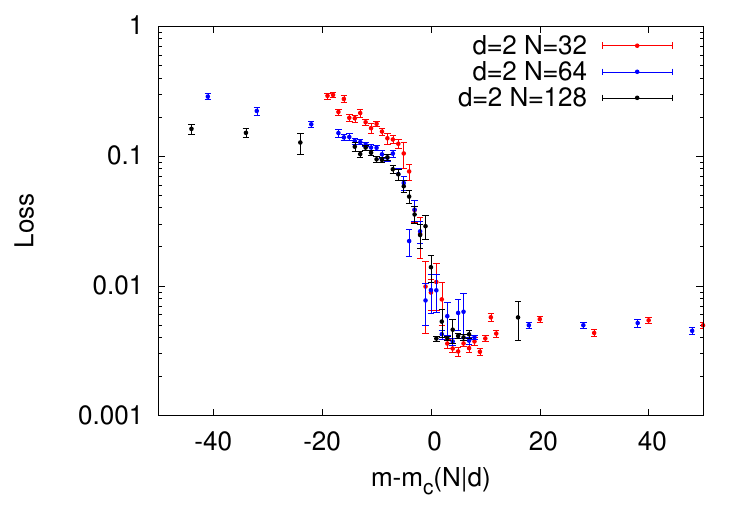}
    \includegraphics[width=0.49\textwidth]{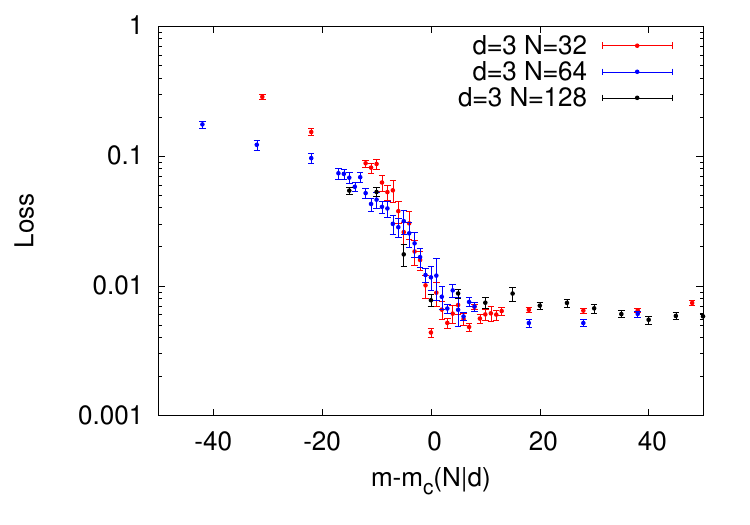}
    \includegraphics[width=0.49\textwidth]{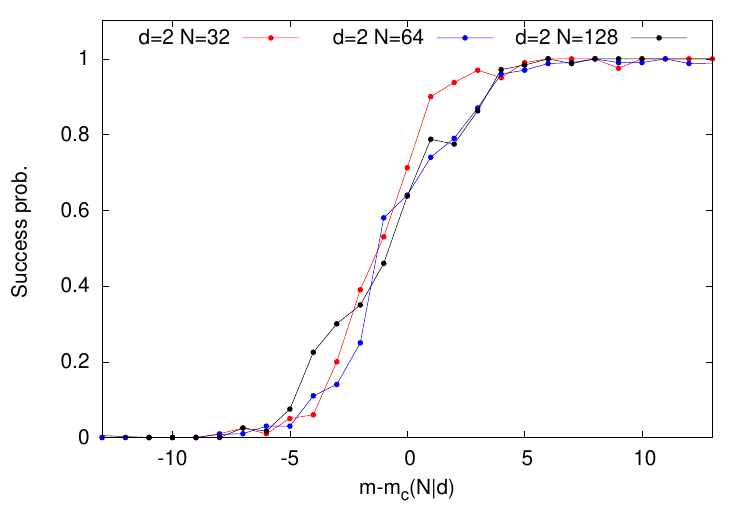}
    \includegraphics[width=0.49\textwidth]{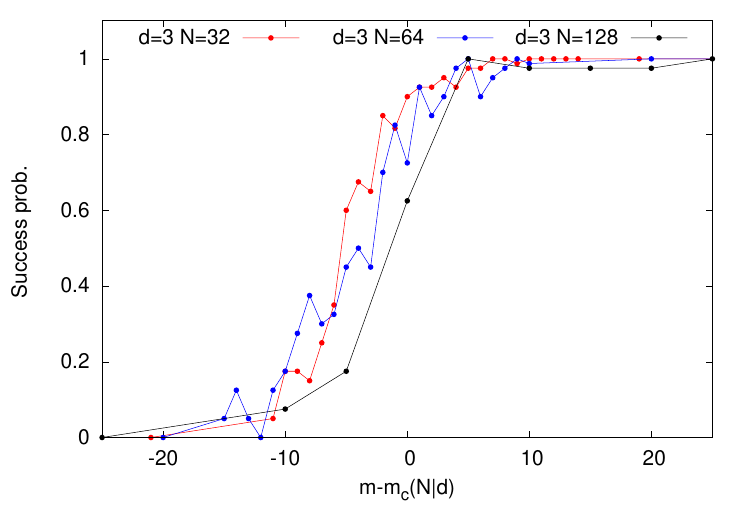}
    \caption{\textbf{Top} panels displays the averaged validation loss as a function of the mini-batch size $m$ for different values of $N$ at $d=2$ (\textbf{left}) and $d=3$ (\textbf{right}). Error bars are the standard deviation of the mean, obtained from $100$ samples. \textbf{Bottom} panels display the fraction of successful learning processes, defined as those reaching a loss function value smaller than the threshold value $0.02$ after $1024$ Monte Carlo Sweeps.}
    \label{fig:loss_reg_1l}
\end{figure}

As a first analysis, we assume that the teacher provides the complete topology to the student. In other words, the student knows the activation function $\phi$, and the position of the non-zero elements in $\mathbf{W}^{*}$, and just needs to infer the sign of the weights in her own neural network
\begin{equation}
\label{mod3eq_student}
    \hat{f}(\vec{x}, \Theta) = \frac{1}{N} \sum_{r=1}^{N}\phi \left( \frac{(\mathbf{W}^\intercal \vec{x})_r}{\sqrt{d}} \right),
\end{equation}
where $\hat{f}(\vec{x}, \Theta)$ is a scalar function, i.e., $\hat{f}: \mathbb{R}^{N} \to \mathbb{R}$, and the set $\Theta$ contains all  the non-zero parameters of the matrix $\mathbf{W}$. Before the learning process starts the non-zero elements of this matrix are set to $\pm 1$ randomly.
Also in this case, the natural choice of the student the learn the matrix is to use the greedy algorithm, with a loss function defined in \eqref{eq::loss}.

Although, the model described by \eqref{mod3eq} and \eqref{mod3eq_student} is simpler with respect to the ones described in Sec.~\ref{mod1} and Sec.~\ref{mod2}, the inference of $\mathbf{W}$ is more complicated since the output layer is just performing an average and this introduces new permutation symmetry, which in turn implies that several permuted weights are able to generalize well the teacher model. For this reason, we are going to study the learning process by just measuring the loss function.

Also in this case, we perform a numerical analysis to understand whether changing the mini-batch size $m$ the learning process undergoes a transition between two phases, one where the inference is impossible and one where the inference is possible. In this case, it is more appropriate to say that the transition is between two phases, one where the student is not able to generalize the teacher model, and one where the generalization is possible. We define a set of pairs $(\vec{x}_{\eta},y_{\eta})_{\eta=1,..,1024}$ to be the validation data set.

In Fig.~\ref{fig:loss_reg_1l}, we present the analysis performed over different values of $N$ and $d$. More precisely, in the top panels we present the averaged validation loss as a function of the hyper-parameter $m$, for two values of $d$, i.e. $d=2$ top left panel and $d=3$ top right panel, and different values of $N=32,\, 64,\, 128$.
For each data set, we observe a clear sharp decrease in the mean value of the loss function when $m$ crosses a critical value $m_c(N|d)$, thus suggesting a sort of first-order transition in the minimization of the loss.
The validation loss function, however, does not reach the $0$ value. This effect is in agreement with the observation written above. This effect is also observed in the Hamming distance, which is not plotted. In this case, we observe that for both values of $d$, the Hamming distance starts from a random guess value for small values of $m$, and reaches at $m_c(N|d)$ a value of $0.15$ for each value of $N$ analyzed. In contrast to the validation loss function that remains constant above the value of $m_c(N|d)$, the Hamming distance increases as well as $m$ increases, showing, therefore, that the configurations found are \textit{far away} from the true planted model, but they are able to generalize well the teacher model.   

Given the jump-like behaviour of the loss shown in the upper panels of Fig.~\ref{fig:loss_reg_1l}, we can easily define a threshold value for the loss that discriminates successful learning processes. This fraction is plotted in the bottom panels of Fig.~\ref{fig:loss_reg_1l}, where we observe good scaling in the success probability of the learning process.

\begin{figure}
    \centering
    \includegraphics[width=0.7\textwidth]{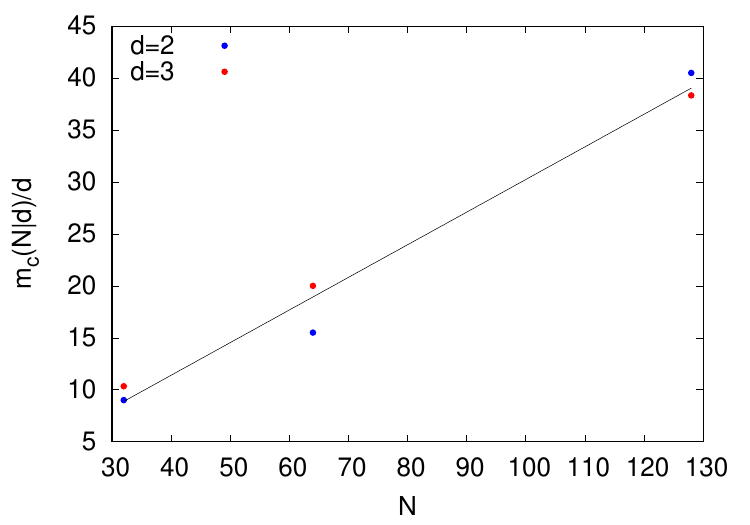}
    \caption{Data for $m_c(N|d)/d$ grow linearly with $N$, thus supporting the scaling $m_c(N|d) \sim d\,N$ for the critical batch size.}
    \label{fig:scalingN1}
\end{figure}

The values for $m_c(N|d) / d$ are reported in Fig.~\ref{fig:scalingN1} as a function of $N$. The linear growth supports the scaling $m_c(N|d) \sim d\,N$ for the critical batch size. In analogy with the argument presented at the end of Sec.~\ref{mod1}, we can compare this critical mini-batch size with the number of parameters to be inferred. In the present model, the parameters in the $\mathbf{W}$ matrix are $dN$ and, given $y$ is a scalar, each mini-batch provides $O(m)$ informative numbers. So one would expect the optimal scaling to be $m_c \sim d N$ and this is confirmed by the data in Fig.~\ref{fig:scalingN1}.

\subsection{Sparse teacher, dense student, one matrix, mismatch-case}
\label{mod4}

For this analysis, we assume that the teacher model is the same as the one described in Sec.~\ref{mod3}. In other words, the components of $\vec{x}\in \mathbb{R}^{N}$ are i.i.d. from $\mathcal{N}(0,1)$ and $y \in \mathbb{R}$ is defined by Eq.~\eqref{mod3eq} with the same activation function $\phi$. In contrast to Sec.~\ref{mod3}, here we assume the student has no knowledge about the topology of the teacher neural network, i.e.\ the non-zero elements of $\mathbf{W}$, while she knows the functional form in Eq.~\eqref{mod3eq}. Because of her lack of knowledge, the best the student can do is to try to model the teacher neural network using a dense matrix $\mathbf{W}$ with real elements.
The learning of the optimal $\mathbf{W}$ matrix can be possibly achieved by optimizing the loss function defined in Eq.~\eqref{eq::loss}. This training process is very similar to the one which is commonly used to train modern neural networks through a deep learning approach, so relevant in many different disciplines. 

Having no knowledge about the teacher neural network, the problem is not anymore an inference problem, but a learning problem. The student wants to find a configuration of the weights that is able to generalize the teacher model. 
Again, our main focus is to understand whether the mini-batch size plays a crucial role in this process of generalizing the teacher model. 
In this particular case, however, a new hyper-parameter comes into play. It is the learning rate $\lambda$. It determines the step size at each iteration while moving toward a minimum of the loss function. It needs to be set \textit{a priori} and can eventually be changed dynamically during the learning process. For the sake of simplicity, we always keep it fixed during the learning process.

For the teacher model, we use a sparse matrix $\mathbf{W}^*$ with $d_t=2$ non-zero elements in each column and in each row: they are set in random positions to $\pm1$ with equal probability.  We use $N=32,\, 64$ and build a  student network with a hidden layer composed of $N$ neurons. The matrix $\mathbf{W}$ is dense, and each element of the matrix is initialized uniformly random in $[0,1]$. We define a set of pairs $(\vec{x}_{\eta},y_{\eta})_{\eta=1,..,1024}$ to be the validation data set.

\begin{figure}
    \centering
    \includegraphics[width=0.48\textwidth]{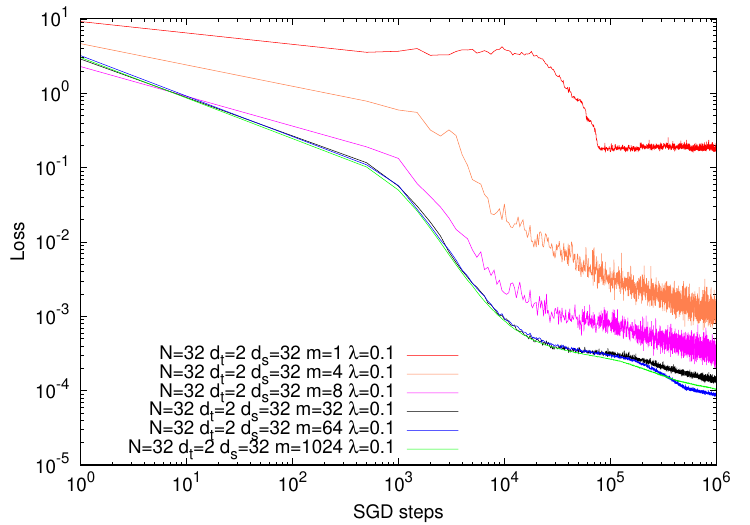}
    \includegraphics[width=0.51\textwidth]{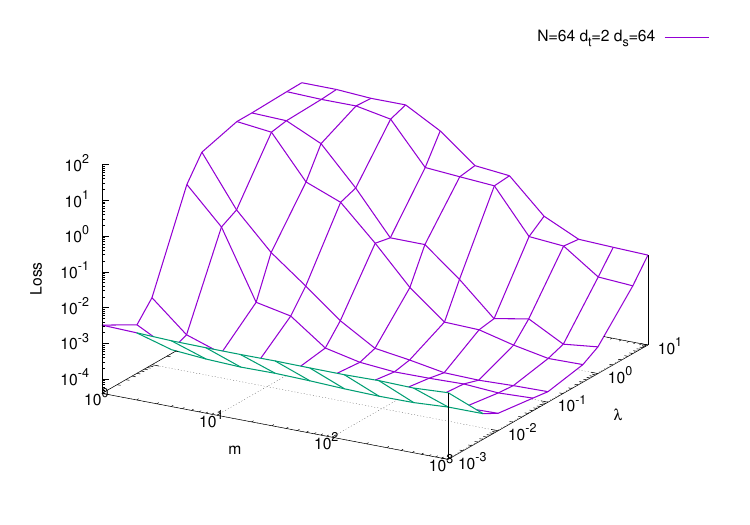}
    \caption{\textbf{Left}: The figure displays the validation loss as a function of the SGD steps (i.e.\ the number of times that we compute the gradient) for the training of a student network with $N=32$ neurons in the hidden layer ($\lambda = 0.1$, $d_t=2$, $d_s=32$) and different values of $m$. \textbf{Right}:  The figure displays the phase diagram of the validation loss achieved after $10^6$ SGD steps as a function of the learning rate $\lambda$ and the mini-batch size $m$ (here $N=64,\,d_t=2,\,d_s=64$).}
    \label{fig:loss_reg_2l}
\end{figure}

In the left panel of Fig.~\ref{fig:loss_reg_2l} we show the evolution of the validation loss as a function of the number of steps of the SGD algorithm (i.e.\ the number of times that we compute the gradient). Data are for a student network with $N=32$ neurons in the hidden layer, and the SGD algorithm uses $\lambda = 0.1$ and different values of $m$. The behaviour of the learning process has a marked dependence on the mini-batch size $m$. For $m=1$ we observe that the SGD dynamics get trapped by local minima with a very large loss function. For larger $m$ values we observe a steady improvement of the generalization, which becomes better both by increasing the SGD steps and the $m$ value. Focusing on the validation loss value at the largest time ($10^6$ SGD steps) we observe that the improvement stops around the optimal value of the mini-batch $m_c \simeq 64$. Above such a value we do not observe any improvement in the generalization (actually the validation loss can even become a little bit larger by further increasing $m$).

For the optimal value $m=64$, it becomes very clear the typical shape of the SGD relaxation, that proceeds through successive plateaus. This is a very well known phenomenon \cite{saad1995line,veiga2022phase,berthier2023learning} and it is interesting to notice that it shows up only if the mini-batch size is set to the optimal value $m_c$.

The behaviour discussed above, where the learning and generalization processes get improved by increasing the mini-batch size $m$, looks very general. In the right panel of Fig.~\ref{fig:loss_reg_2l} we show the value of the validation loss achieved after $10^6$ SGD steps as a function of the learning rate $\lambda$ and the mini-batch size $m$. We observe that for any value of $\lambda$ the generalization error decreases as a function of $m$ until an optimal value $m_c$ is reached. The value of $m_c$ depends on the learning rate $\lambda$ but its existence seems very robust in a wide $\lambda$ range. It is worth noticing that the validation loss varies by several orders of magnitude increasing the value of $m$.

\section{Discussion}
\label{Conc}

Given the central role played by the mini-batch in training artificial neural networks, it is surprising that very little results are available on the effects of changing its size. In this work we try to fill this gap by presenting an accurate numerical analysis that better quantifies the role of the mini-batch size $m$ in training two-layer artificial neural networks. Working within the teacher-student scenario and fixing the teacher to be a sparse neural network, we study four different models for the student neural network. In all the cases we observe a crucial dependence of the generalization error on the mini-batch size $m$. In some cases, such a strong dependence turns into a sharp phase transition between phases where the student is able or unable to generalize well the teacher's neural network. In other words, we observed that above the critical value $m_c$, the student can either infer exactly the teacher model or at least generalize it very well.

The phase transitions we observed are closely tied to the model's architecture, particularly the makeup of the output layer. Models designed for classification tasks, with just one neuron in the output layer, demonstrated distinct behavior compared to those geared for regression tasks with multiple neurons. The presence of a classification task introduces new permutations, adding complexity to the generalization process. This highlights the critical role of both the output layer's structure and the mini-batch size, especially in models where outputs are simple averages of inputs, leading to new permutation symmetries.

The robustness of our results is supported by the fact we find similar behavior in four different models, where the kind of task and the amount of information provided to the student vary. In particular we have studied both the case where the network topology is known to the students and the case where it is not. Moreover we have studied a regression problem where the output of network has the same input size and a simpler classification problem where the network has a single output neuron. To train the student network we have used SGD (in case the network parameters are real variables) and a SGD-like algorithm working with discrete variables otherwise. In all cases the effect of changing the mini-batch size is clear.

To extend our findings to various architectures and optimizers, one could embark on similar numerical experiments but in a more intricate setting. Firstly, the sparsity of the model is an integral aspect that must be considered when attempting to generalize results. Moreover, it's imperative that the teacher model remains encapsulated within the student model, ensuring consistent and meaningful extrapolation of outcomes. Additionally, selecting a deeper architecture, choosing the appropriate dataset, and committing the necessary computational resources for training the neural network with varying mini-batch sizes is essential. When considering optimizers, besides traditional methods (as the one used in this manuscript), there's potential in employing Simulated Annealing (SA) \cite{kirkpatrick1983optimization}, or Parallel Tempering \cite{swendsen1986replica}, or Monte Carlo algorithms \cite{metropolis1949monte} with a specific temperature setting, akin to the approach we adopted in \cite{angelini2023stochastic}. 

The general picture that comes out from this study is the following. For small values of $m$ the information provided to the algorithm at each step is too noisy and the training process get stuck in sub-optimal configurations, that are called \emph{glassy states} in statistical physics. Only for $m$ large enough the information collected at each step from the data allows the training algorithm to optimize the loss function and in turn produce a student network with good generalization skills.

We have shown that in some models these two regimes are separated by a clear phase transition, and the critical value $m_c$ scales with the model parameters (e.g.\ the input size $N$ and the teacher connectivity $d$) in a way to match the amount information provided in each batch with the number of parameters to be assigned/inferred in the training process. This matching is a very simple rule that may help in understanding \emph{a priori} what is the optimal value for the mini-batch size.

\section*{Data availability statement}
The numerical codes used in this study and the data that support the findings are available from the corresponding author upon request.

\bibliography{sample}

\section*{Acknowledgements}

We thank Jean Barbier and Manuel Saenz for useful discussions. FM and RM thank  the FARE grant No. R167TEEE7B and  Simons Foundation Grant No. 454949. RM work is also supported by \#NEXTGENERATIONEU (NGEU) and funded by the Ministry of University and Research (MUR), National Recovery and Resilience Plan (NRRP), project MNESYS (PE0000006) "A Multiscale integrated approach to the study of the nervous system in health and disease" (DN. 1553 11.10.2022). We acknowledge financial support from PNRR MUR project PE0000013-FAIR.

\section*{Author contributions statement}

Both authors contributed equally to all aspects of this work, but the numerical simulation which have been run mainly by R.~Marino.

\end{document}